# THE CONTROL OF A BEAMLINE OVER INTRANET

X.J. Yu, Q. P. Wang, P. S. Xu

University of Science and Technology of China, Hefei, Anhui 230029, China,

## Abstract

The machines and beamlines controlled by VME industrial networks are very popular in accelerator faculties. Recently new software technology, among of which are Internet/Intranet application, Java language, and distributed calculating environment, changes the control manner rapidly. A program based on DCOM[1] is composed to control of a variable included angle spherical grating monochromator beamline at National Synchrotron Radiation Laboratory (NSRL) in China. The control computer with a residential DCOM program is connected to Intranet by LAN, over which the user-end-operating program located in another computer sends driving beamline units' commands to the control computer. And also a web page coded in Java, published by the WWW service running in the control computer, is simply illustrated how to use web browser to query the states of or to control the beamline units.

## 1 INTRODUCTION

VME bus system, widely used in synchrotron radiation facilities [1,2], is very reliable, ease of integration and large commercial availability. Nevertheless its disadvantage seems to be apparent, the VME system executive unit is less powerful than PC unit and usually not compatible with PC software, many advantage features based on Windows, for instance, ActiveX,

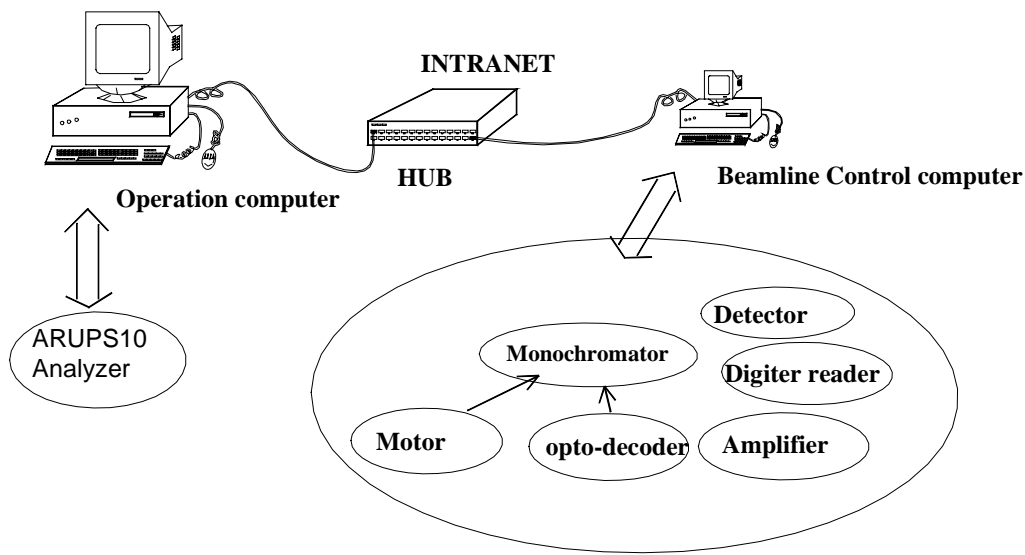

Figure 1: The distributed control logic of a beamline at NSRL

DCOM, et al, can not be used in the VME system. As the Windows technology advanced, today the Windows operation system has reached a summit of stability and extendibility. It is no surprise that a W2K server can run a few months without breaking down. Moreover, a vast of visual developing tools, giant reservoir of sample codes across Internet on the base of PC and Windows system, can give a great convenience in the development of beamline control software. A beamline control system comprised of PC units interconnected by Intranet is introduced to control a variable included angle grating monochromator beamline that has been described elsewhere [3] in NSRL.

---

[1] http://www.microsoft.com/TechNet/winnt/Winntas/technote/dcomtowp.asp

The control logic is made up two parts, one is an ActiveX server controlling the hardware, the other is user operating interface which is distributed on Intranet/Internet can be a standalone program coded in C/C++ or WEB based page coded in JAVA.

## 2 STANDALONE PROGRAM

This standalone program has been discussed in detail [4]. An ActiveX server running on a beamline control computer takes control of a set of hardware units of a beamline, which may be made up of motor drivers, opto-decoders, amplifier, digital readers, detectors et al, Fig.1. As we have known, for the calculating logic of a variable angle grating monochromator is quite complicated, the spent time in determining the optics' positions for a single photon energy by the program based on a VME bus system has been reported up to one third second [2]. It is not applicable for that it will integrate to large amount of time during a photon energy scan with a step on monochromator resolution level. So it is adopted a cubic curve to fit the positions of mirror and grating against photon energy. But the drawback is that it can not change the monochromator parameters in real time, while this is very important during the phase of beamline commissioning. In our case it needs only several milliseconds to do a real time calculation for a fast PC computer.

The server side program loading steps are as following. At first, as the client side program calling to the server interface, it invokes the load of the ActiveX server in which one public class initializes a public variable that represents the instance of one set of the beamline hardware units. Any function sequentially operating the hardware unit must be accessed from this instance variable. And in the consequently call to the ActiveX does not re-initialize the variable again so that it keeps tracing the current states of the hardware units.

In the server side program, every beamline unit is related to a form window, which is usually created invisibly, because when ActiveX runs in a remote computer no parameters needed to be entered. On which some Windows standard controls, such as serial port control, timer control, whose properties can be preset or changed by calling to interface functions supplied by the ActiveX server, can be placed. All modules must be dedicatedly designed to process every encountered event of the hardware units to prevent the program crash on any critical error. In ActiveX server no massage box can be brought up because no one can dismiss it at the remote computer.

A client on an operation computer establishes a communication with the ActiveX server in the beamline computer by way of a HUB, from which is very convenient to place the client computer anywhere. One can built the client by using Borland C++ builder or other visual programming tools. There are two types of connection between the client and the server, dynamic and static. DCOM is usually a dynamic one. When using dynamic connection, a DCOM component is created directly by CoCreateInstanceEx function whose ComputerName parameter is set to the remoter beamline computer name. The DCOM component acts as an agent to pass any hardware call to the remote interface and we can regardless of what happens in the intermediate process. The dynamic connection between the client and the DCOM server does degrade the whole performance of the client software. Another solution that is using static connection can improve the performance dramatically. It is required to copy the ActiveX executable file to and registered in the client computer, then we can import it to the client side program to create a package component. So a reference instance can be obtained by pulling it to a designing frame window, several files wrapping all interface functions of the ActiveX server are automatically created. The RemoteMachineName property of which is also set to the remote computer name. The created files are then compiled and linked with the other client modules to form a static connection client side executable program which would be faster and more stable than the dynamic connection client.

## 3 JAVA PROGRAMMING

The above client side program is standalone software, the other way of control is no client side program but a web page residing in a web server which is written by Java language. Java language is very common in web design that is platform independent. If we write beamline control software in Java language, then we can control and access the actions of all units of a beamline regardless anywhere you are. Before you can access the beamline hardware from a WEB page, at first, an IIS Internet information server must be set up. Second, a component that enables DCOM being accessed across http protocol must be installed, which makes it possible to call DCOM interface from a web page. Finally a Java program wrapping interface which calls the ActiveX server same as that has been discussed above with some graphically input panels and buttons is developed and then deployed in the IIS server.

## 4 SUMMARY

A control software built on base of Intranet and made up of PC computers is realized by us. The challenge is encountered in design the features of ActiveX server and many key problems have been overcome. The beamline control over Internet WEB page is a representation of the trend of thin client computer. And also the activities of the beamline are easily inspected by people in the world.

## REFERENCES

[1] J. Krempasky, et al., "A system for controlling the variable angle spherical grating monochromators at Elettra", SPIE, 3150 (1997) 76-85
[2] David E. Eisert, et al., "Computer control of the SRC high-resolution beamline", Rev. Sci. Instrum. 66(2) (1995) 1671-1673
[3] X.J. Yu et al., "Optimization of a variable angle spherical grating monochromator", Nucl. Inst. And Meth. A 467-8 (2001), 597-600
[4] X.J. Yu et al., "Computer control of surface physics beamline at NSRL", Nuclear Techniques (in Chinese), 24 (2001), 534-539